\documentclass[fleqn,usenatbib]{mnras}

\usepackage{newtxtext,newtxmath}

\usepackage[T1]{fontenc}
\usepackage{ae,aecompl}

\usepackage{graphicx}	
\usepackage{amsmath}	

\title[LOFAR observations of SGRB 181123B]{\textit{LOFAR early-time search for coherent radio emission from Short GRB 181123B}}

\author[Rowlinson et al.]{A. Rowlinson$^{1,2}$\thanks{Contact e-mail: \href{mailto:b.a.rowlinson@uva.nl}{b.a.rowlinson@uva.nl}}, R.L.C. Starling$^{3}$, K. Gourdji$^{1}$, G. E. Anderson$^{4}$, S. ter Veen$^{2}$, \and S. Mandhai$^{3}$, R.A.M.J. Wijers$^{1}$, T.W. Shimwell$^{2}$, A.J. van der Horst$^{5,6}$
\\
$^{1}$ Anton Pannekoek Institute, University of Amsterdam, Postbus 94249, 1090 GE Amsterdam, The Netherlands\\
$^{2}$ ASTRON, the Netherlands Institute for Radio Astronomy, Oude Hoogeveensedijk 4, 7991 PD, Dwingeloo, The Netherlands\\
$^{3}$ School of Physics and Astronomy, University of Leicester, University Road, Leicester LE1 7RH, UK\\
$^{4}$ International Centre for Radio Astronomy Research, Curtin University, GPO Box U1987, Perth, WA 6845, Australia\\
$^{5}$ Department of Physics, The George Washington University, 725 21st Street NW, Washington, DC 20052, USA\\
$^{6}$ Astronomy, Physics, and Statistics Institute of Sciences (APSIS), 725 21st Street NW, Washington, DC 20052, USA}

\date{Last updated 2015 May 22; in original form 2013 September 5}

\pubyear{2020}

\begin{document}
\label{firstpage}
\pagerange{\pageref{firstpage}--\pageref{lastpage}}
\maketitle

\begin{abstract}
The mergers of two neutron stars are typically accompanied by broad-band electromagnetic emission from either a relativistic jet or a kilonova. It has also been long predicted that coherent radio emission will occur during the merger phase or from a newly formed neutron star remnant, however this emission has not been seen to date. This paper presents the deepest limits for this emission from a neutron star merger folowing triggered LOFAR observations of the short gamma-ray burst (SGRB) 181123B, starting 4.4 minutes after the GRB occurred. During the X-ray plateau phase, a signature of ongoing energy injection, we detect no radio emission to a 3$\sigma$ limit of 153 mJy at 144 MHz (image integration time of 136 seconds), which is significantly fainter than the predicted emission from a standard neutron star. At a redshift of 1.8, this corresponds to a luminosity of $2.5 \times 10^{44}$ erg s$^{-1}$. Snapshot images were made of the radio observation on a range of timescales, targeting short duration radio flashes similar to fast radio bursts (FRBs). No emission was detected in the snapshot images at the location of GRB 181123B enabling constraints to be placed on the prompt coherent radio emission model and emission predicted to occur when a neutron star collapses to form a black hole. At the putative host redshift of 1.8 for GRB 181123B, the non detection of the prompt radio emission is two orders of magnitude lower than expected for magnetic reconnection models for prompt GRB emission and no magnetar emission is expected.
\end{abstract}




\section{Introduction}
\label{sec:introduction}

The detection and association of the gravitational wave event, GW170817, and the short Gamma-Ray Burst (SGRB) 170817A confirmed the theory that the progenitor of many SGRBs is the merger of two neutron stars \citep{abbott2017}. However, the nature of the remnant formed via this merger is still debated, with the two competing models being a black hole or a massive, rapidly rotating, highly magnetised neutron star \citep[hereafter referred to as a magnetar; e.g.][]{fong2016,ai2018,piro2019,liu2020}. Current gravitational wave observatories lack the sensitivity required to answer this question \citep[e.g.][]{abbott2019}, though the next generations of gravitational wave observatories may be able to measure the properties of the remnant in the future \citep[e.g.][]{banagiri2020}.

Tantalising observational evidence shows that the central engine powering the GRB is active long after the merger of the two neutron stars, leading to flares and plateau phases in the electromagnetic light curve \citep{rowlinson2013}. \cite{rowlinson2013} showed that the plateau phases in X-ray light curves following many SGRBs are consistent with the central engine being a magnetar. While support for this model has increased, there is currently no `smoking gun' observation to prove that a magnetar was formed via the merger of two neutron stars. However, as outlined by \cite{rowlinson2019} and references therein, if a magnetar is formed we would expect it to produce copious amounts of radio emission via a range of mechanisms. This radio emission is not expected if the remnant formed is a black hole. Identifying this radio emission would thus provide convincing support for the magnetar model.

Following the discovery of Fast Radio Bursts \citep[FRBs; e.g.][]{lorimer2007,thornton2013}, several of the progenitor theories suggested they could come from cataclysmic events such as binary neutron star mergers \citep[e.g.][]{zhang2014}. The discovery of repeating FRBs \citep[such as FRB 121102;][]{spitler2014,spitler2016}, showed that at least some FRBs were not coming from cataclysmic events. Therefore, either they are all not due to cataclysmic events or there are at least two different progenitors possible for FRBs. The recent detection of FRBs from the Galactic magentar SGR 1935+21 \citep{andersen2020,kirsten2020} further supports the possibility of bright coherent radio emission coming from newborn magnetars formed during GRBs. Recent advances in the localisation of FRBs within their host galaxies have revealed a variety of FRB host galaxy types and environments. Host galaxy comparison studies have found that a subset of the FRB hosts are consistent with the hosts of SGRBs \citep{margalit2019,li2020}. \cite{gourdji2020} consider the likelihood that some of those non-repeating FRBs are consistent with some of the coherent radio emission models for compact binary mergers.

Previous efforts to identify this coherent radio emission following SGRBs have been unsuccessful. Early searches have either been very insensitive ($>100$ Jy) and/or have only sampled a small number of SGRBs to date \citep[][]{cortiglioni1981,inzani1982,koranyi1995,dessenne1996,balsano1998}. With the advent of the next generation of radio telescopes, with either large fields of view or rapid slew capabilities, searches have resumed in earnest to find this elusive emission. Recent searches using the Murchison Widefield Array \citep[MWA;][]{tingay2013} and the Owens Valley Radio Observatory Long Wavelength Array \citep[OVRO-LWA;][]{hallinan2014} and the first station of the Long Wavelength Array \citep[LWA1;][]{ellingson2013} have started obtaining constraining limits for SGRBs at low radio frequencies \citep[for SGRBs 150424A, 170112A and 180805A;][]{obenberger2014, kaplan2015,anderson2018,rowlinson2019,anderson2020}. Meanwhile the LOw Frequency ARray \citep[LOFAR;][]{vanhaarlem2013} has also demonstrated its potential to obtain deep constraints on this emission by following up the long GRB 180706A \citep{rowlinson2019b}. Additionally, at 1.4 GHz, the Australian Square Kilometer Array Pathfinder \citep[ASKAP;][]{hotan2014} has followed up 20 GRBs (including four SGRBs) with their rapid response system \citep{bouwhuis2020}. After searching their data for FRBs, they concluded there was no pulsed radio emission above 26 Jy ms. Although these are all non-detections to date, they have proven that the required sensitivities can be obtained to test the various models \citep{rowlinson2019}. Many SGRBs do not show evidence of ongoing central engine activity and, hence, are more likely to have formed a black hole remnant so radio emission is not expected. Also, of the SGRBs with ongoing energy injection, these need to either be sufficiently energetic or nearby to produce detectable radio emission \citep{rowlinson2019}.

Due to its sensitivity and rapid response mode enabling observations to start within 5 minutes of an alert, LOFAR is an ideal facility to chase the predicted radio emission. Since 2017, LOFAR has been responding fully automatically to GRB alerts and, on 2018 November 23, was successful in obtaining data following a SGRB detected by the Neil Gehrels Swift Observatory \citep[hereafter referred to as {\it Swift};][]{gehrels2004}. This paper presents the deep search for coherent radio emission following this SGRB. In Section \ref{sec:observations}, we present the observational data obtained for this event, while in Section \ref{sec:interpretation} we compare the observations to predictions tailored to this event.

\begin{figure}
\centering
\includegraphics[width=0.45\textwidth]{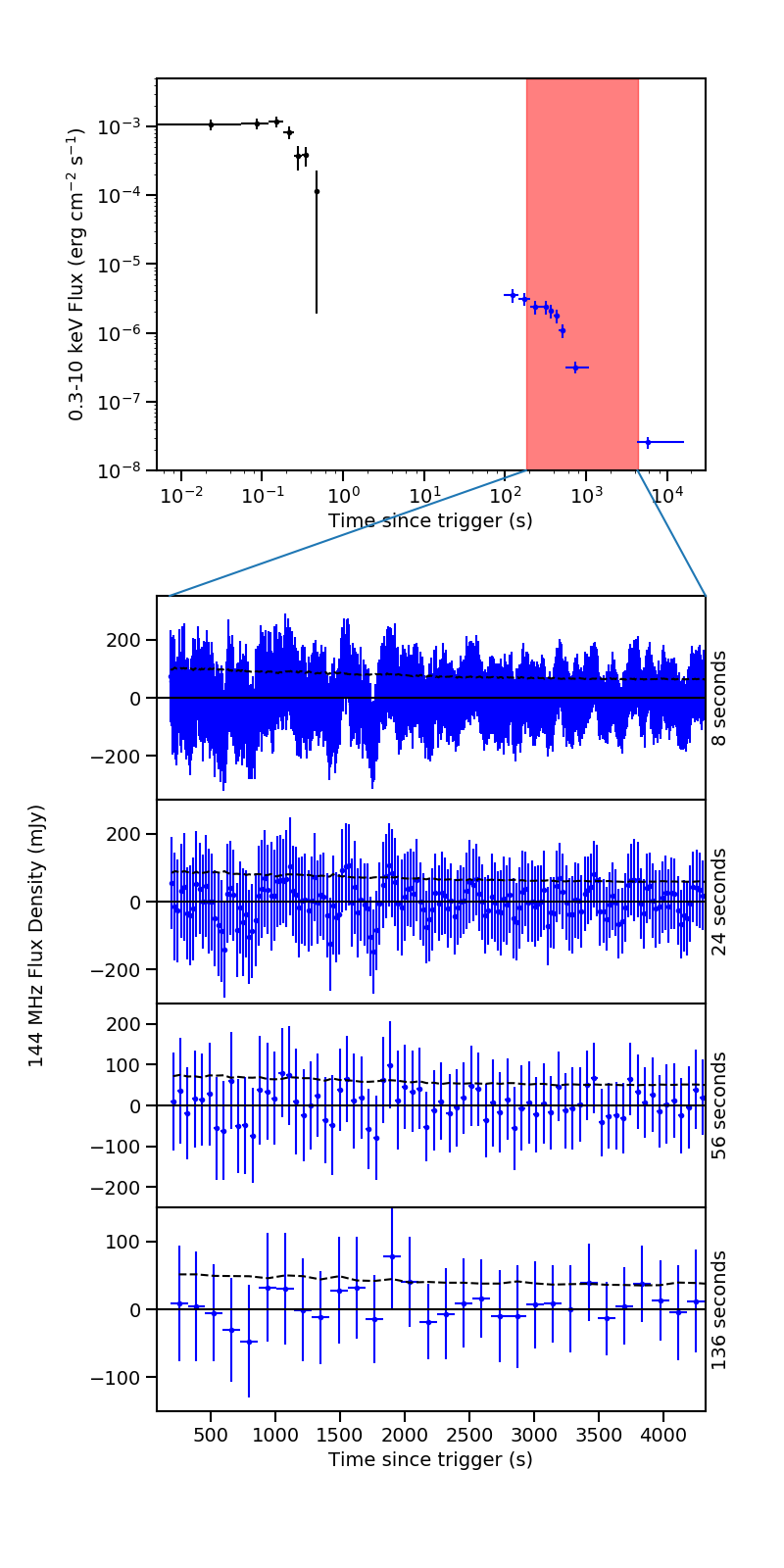}
\caption{This figure, adapted from \citet{rowlinson2019b}, shows the X-ray flux lightcurve of GRB 181123B in the 0.3--10 keV energy band (top panel) in which data from the BAT and XRT are shown with the black and blue data points respectively. The time of the observation obtained by LOFAR is illustrated by the red shaded region. The LOFAR 144MHz flux density limits were measured at the position of GRB 181123B on four snapshot timescales (8, 24, 56 and 136 seconds) and are illustrated in the bottom panel. We show the flux density of 0 Jy with a solid black line. The rms noise of each image is measured in the inner $\frac{1}{8}$th of the image and is plotted with the dashed line.}
\label{fig:LC}
\end{figure}

\section{Observations of GRB 181123B}
\label{sec:observations}

\subsection{Swift Observations}
\label{sec:Swiftobs}

The {\it Swift} Burst Alert Telescope  \citep[BAT;][]{barthelmy2005} triggered and located GRB 181123B (trigger=873186) on 2018 November 23 at 05:33:03\,UT \citep{lien2018}. {\it Swift} slewed immediately to the burst, and X-Ray Telescope \citep[XRT;][]{burrows2005} observations began 80.2 s after the BAT trigger, locating the X-ray afterglow to within a 90\% error region of 1.6$''$ radius at a position of RA: 184.36686 degrees, Dec: 14.59788 degrees\footnote{www.swift.ac.uk/xrt{\_}positions}. The duration of $T_{\rm 90} = 0.4$\,s \citep{lien2018} and the lag analysis \citep{norris2018} confirm that this is a short GRB, possibly accompanied by extended emission (observed at low significance). The GRB was also detected above the BAT energy band by Insight-HXMT, which recorded a duration of 0.23 s \citep{InsightGCN}.
{\it Swift's} UltraViolet and Optical Telescope \citep[UVOT; ][]{roming2005} did not detect a counterpart \citep{oates2018}.

In Figure \ref{fig:LC} we show the {\it Swift} BAT and XRT 0.3--10 keV observed flux light curves (black and blue data points respectively) obtained using the {\it Swift} Burst Analyser \citep{evans2010}. The light curve consists of a single prompt, $\gamma$-ray flare followed by a fading X-ray counterpart that is modelled using a single power law decline of $\alpha_{X} = 1.31^{+0.15}_{-0.14}$ \citep{burrows2018}. 

\subsection{Other Observations}
\label{sec:otherobs}

A faint and likely extended near-infrared counterpart within the XRT error circle was reported 9.2 h after the trigger with $i(AB) \sim 23.32\pm0.25$ \citep{FongGCNGemini} and $J(AB) \sim 22.94\pm0.19$ \citep{PatersonGCN1}. A second observation of that source found a marginal detection with $J(AB)>23.3 $ \citep{PatersonGCN2} at 3.38 days. Following further analysis, this source is identified as the putative host galaxy of GRB 181123B at a redshift of $z = 1.8$ with a chance alignment probability of $\lesssim 0.44$\% \citep{paterson2020}. We note that only one emission line was detected in the optical spectrum. This emission line is attributed to H$\beta$ resulting in the redshift being 1.8, which is consistent with the photometric constraints \citep{paterson2020}.

An 8.3 hr radio observation was also obtained using the Australia Telescope Compact Array rapid-response mode, which automatically triggered observations when the source was above the horizon, starting at 12.6 hrs post-burst and providing 3 sigma upper limits of 34 and 32 microJy/beam, respectively \citep{anderson2020}.

\subsection{LOFAR Observations}
\label{sec:LOFARobs}

\begin{figure}
\centering
\includegraphics[width=0.48\textwidth]{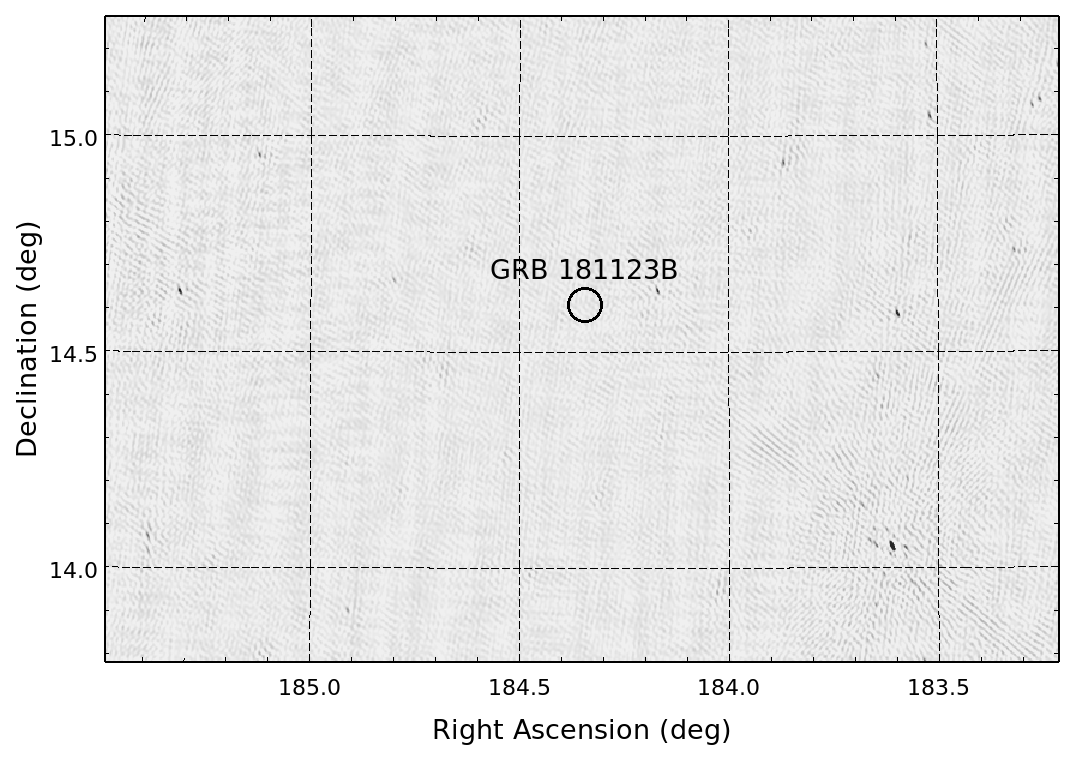}
\caption{This image of the region surrounding GRB 181123B was attained using 71.2 minutes of LOFAR data. The circle shows the location of GRB 181123B and the 3$\sigma$ upper limit on the flux density of this event is 60 mJy beam$^{-1}$.}
\label{fig:radioImg}
\end{figure}

Since November 2017, LOFAR has been able to fully automatically respond to transient alerts, which are typically communicated via VOEvents \citep{williams2006}. We utilise VOEvents that are redistributed via the {\sc 4 Pi Sky broker} \citep[][]{staley2016} and receive them using the {\sc Comet} broker software \citep{swinbank2014}. Transients are then filtered according to predetermined triggering criteria, including: identification of source (GRB), {\it Swift} trigger integration time ($\le$ 1 second), elevation of the source ($\ge$15 degrees) and calibrator availability. Following this, an {\sc xml} observing request is sent to the LOFAR system. GRB 181123B passed the triggering criteria and LOFAR observations started 4.4 minutes after the GRB occurred. A 2 hour LOFAR observation was started at 05:37:25 UTC on 2018 November 23 and was centred on the BAT localisation of GRB 181123B. The observation time is highlighted by the red shaded region in Figure \ref{fig:LC}. Unfortunately, due to a scheduling error, the full observation was not completed (total usable observation time attained: 71.2 minutes) and the calibrator observation was not completed automatically. The LOFAR Radio Observatory manually scheduled a 15 minute calibrator observation of 3C286 at 13:20:03 UTC on 2018 November 23.

In order to benefit from the deep 8 hour images attained for much of the Northern Hemisphere by the LOFAR Two Meter Sky Survey \citep[LoTSS;][]{shimwell2018}, we matched our observational setup to that of LoTSS. When available, the LoTSS observations provide a deep comparison image and accurate sky model of the field for calibration purposes. For these observations we used the LOFAR High Band Antennas (HBA) with a central frequency of 144 MHz. The frequency range is 120--168 MHz, covered by 244 sub-bands each with a bandwidth of 195.3 kHz. We used the Dutch LOFAR stations, 23 core stations and 11 remote stations. The data were recorded using a time-step of 1 second and 64 channels per sub-band. Our observations were pre-processed using the standard methods for LOFAR \citep{vanhaarlem2013}.

\subsubsection{Calibration}
\label{sec:calibration}

Following the method outlined in \cite{rowlinson2019b}, the LOFAR observations were calibrated using {\sc prefactor}\footnote{\url{https://github.com/lofar-astron/prefactor}} and a strategy based upon that presented in \cite{vanweeren2016}. Both the target and calibrator observations were flagged for excess radio frequency interference using {\sc AOFlagger} \citep{offringa2010, offringa2012}. One of the brightest sources in the radio sky, Virgo A, is 3.9 degrees from the position of the GRB and dominated the radio image. Using the detailed skymodel provided with {\sc prefactor}, we removed the contribution of Virgo A from the observations. The calibrator and target visibility data were averaged in time to 8 seconds and in frequency to 48.82 kHz (4 channels per subband).

Using the model obtained by \cite{scaife2012}, we obtained the diagonal gain solutions for the calibrator source 3C286, which were then transferred to the target visibility data. The target subbands were combined in groups of 10, resulting in combined datasets of 1.953 MHz. A sky model of the target field was obtained using the global sky model developed by \cite{scheers2011} and the TIFR GMRT Sky Survey at 150 MHz \citep[TGSS; ][]{intema2017}\footnote{url{http://tgssadr.strw.leidenuniv.nl/doku.php}}. The sky model of the field was then used to conduct a phase calibration of the target visibilities.

\subsubsection{Imaging}
\label{sec:Imaging}
We imaged the full LOFAR observation using {\sc WSClean} \citep{offringa2014} using a primary beam correction, Briggs weighting (robustness of $-0.5$), a pixel scale of 10 arcseconds and baselines up to 12km. Cleaning was conducted using an automatic threshold and $10^4$ iterations. The final image has a central frequency of 144 MHz and a bandwidth of 48 MHz. The image has a typical angular resolution of $\sim$30 arcsec. The region surrounding GRB 181123B is shown in Figure \ref{fig:radioImg} and the image RMS at the GRB location (30 arcsecond radius) is 20 mJy beam$^{-1}$, corresponding to a 3$\sigma$ upper limit of 60 mJy beam$^{-1}$. Using the Python Source Extractor \citep[{\sc PySE};][]{carbone2018} we also conduct a forced source extraction at the position of the GRB holding the shape and size of the Gaussian shape fitted fixed to the restoring beam shape. We measure a peak flux density of $-29 \pm 38$ mJy beam$^{-1}$ \citep[the uncertainty on this value is as measured by {\sc PySE};][]{carbone2018}.

We created a Stokes I image of the visibilities using {\sc WSClean} \citep{offringa2014}\footnote{http://wsclean.sourceforge.net} with Briggs weighting, a pixel scale of 10 arcseconds and baselines up to 12 km. As the image integration times increased, we found that the automatic {\sc Clean} process in WSClean was diverging, likely due to remaining noise contributions following the subtraction of Virgo A and confusion noise, so the {\sc Clean} process was stopped after 10,000 iterations. Therefore the typical rms noise is not expected to follow the expected relationship that the rms noise drops as $t^{\frac{1}{2}}$, where $t$ is the integration time. 

As the X-ray data may show a plateau phase out to $\sim$400 seconds, we first created a radio image using the first 136 seconds of data to search for emission associated with this phase. The image RMS at the GRB location (30 arcsecond radius) is 51 mJy beam$^{-1}$, corresponding to a 3$\sigma$ upper limit of 153 mJy beam$^{-1}$. At the putative host galaxy redshift of 1.8, this corresponds to a luminosity of $2.5 \times 10^{44}$ erg s$^{-1}$. Using {\sc PySE}, we measure a peak flux density of $-9\pm50$ mJy beam$^{-1}$, corresponding to a non-detection. Additionally, using the intervals-out option in {\sc WSClean}, we created snapshot images using the source-subtracted visibilities of durations 8, 24, 56, 136 seconds (the motivation for this range of time scales is outlined in Section \ref{sec:FRBs}).  

We use the monitoring list capability of the LOFAR Transients Pipeline \citep[{\sc TraP}; ][]{swinbank2015} to measure the flux density at the location of the GRB. {\sc TraP} also monitors the typical rms noise in the inner $\frac{1}{8}$th of the input images. In Table \ref{table:rms}, we give the typical rms noise for the different imaging time scales. 

\begin{table}
\centering
\begin{tabular}{|c c|} 
\hline
Time scale & rms noise \\
(seconds)  & (mJy beam$^{-1}$) \\
\hline
8           & $76\pm12 $ \\
24          & $69\pm9  $ \\
56          & $58\pm8  $ \\
138         & $42\pm5  $ \\
\hline
\end{tabular}
\caption{Here we provide the average rms noise and its 1$\sigma$ standard deviation for each set of images from the four time scales considered.}
\label{table:rms}
\end{table}

The light curves were obtained from {\sc TraP} for each of the four time scales used and are plotted as the blue data points in Figure \ref{fig:LC}. We show the image rms using the black dashed line and it can be seen that GRB flux densities are consistent with the image rms noise (note, when the local rms is lower than the image rms, the blue data points can lie below the black dashed lines). Therefore, no coherent emission was detected from GRB 181123B in this analysis.

\section{Modelling and interpretation}
\label{sec:interpretation}

In the following analysis, we need to utilise the rest frame properties of GRB 181123B in order to compare our observations to the various models predicting coherent radio emission. As outlined in Section \ref{sec:otherobs}, there is a host galaxy candidate at a redshift of $z \sim 1.8$ with a very low probability of chance alignment and we consider the implications of this on each model. However, as the redshift is not from a direct measurement of the afterglow and hence still has the possibility of being a chance association, we also treat the redshift as unknown in the following analysis. In Section \ref{sec:promptEmission}, we calculate the minimum redshift ($z\sim1.05$) above which the prompt radio emission can be probed by our LOFAR observations assuming dispersion effects. In Section \ref{sec:magnetarModel}, we use an average SGRB redshift of 0.7 to fit the magnetar model \citep[consistent with analysis in ][]{rowlinson2013, rowlinson2019, rowlinson2019b,anderson2020b} and then analytically scale the results to a range of redshifts.

\subsection{Propagation effects for low frequency radio emission}
\label{sec:propagationAffects}

Coherent radio emission is known to be subject to significant propagation affects, limiting its ability to be detectable at low radio frequencies. Plasma close to the source is opaque below a fixed frequency that is directly proportional to the number density of electrons in the plasma, thus dense regions may be able to block coherent radio emission at LOFAR frequencies. \cite{zhang2014} showed that in the case of SGRBs, such as GRB 181123B, the emission is expected to be able to escape along the jet propagation axis.

Additionally, the surrounding medium can interact with the low frequency photons, leading to free-free absorption, with a strong dependence on the temperature and density of the surrounding interstellar or intergalactic medium. Using the X-ray spectrum of GRB 181123B, we are able to estimate the absorption due to neutral hydrogen along the line of sight to this GRB. Using the automated X-ray spectrum provided by the UK {\it Swift} Science Data Centre, GRB 181123B has an intrinsic absorption column of $N_{H} = 2.5^{+6.6}_{-2.5} \times 10^{20}$ cm$^{-2}$. This is a low absorption column, consistent with zero intrinsic absorption, showing that GRB 181123B most likely occurred in a very low density medium so free-free absorption is expected to be low.

These propagation affects are considered in depth by \cite{rowlinson2019}, who show that they are likely to not affect the coherent radio emission in compact binary mergers such as the likely progenitor of GRB 181123B.

\subsection{Constraints on prompt emission}
\label{sec:promptEmission}

We are able to place constraints on the presence of prompt radio emission from GRB 181123B even though the LOFAR observations do not cover the same time period as the prompt gamma-ray emission. This is because it takes longer for radio emission to propagate to the Earth than it does for the gamma-ray emission. This is due to dispersion delay, a frequency dependent delay due to the the integrated column density of free electrons along the line of sight. The dispersion delay, $\tau$, in seconds is given by
\begin{eqnarray}
\tau = \frac{\rm DM}{241 \nu_{\rm GHz}^{2}} s, \label{eqn:disp_delay}
                                  \end{eqnarray}
where DM is the dispersion measure in pc cm$^{-3}$ and $\nu_{\rm GHz}$ is the observing frequency in GHz \citep{taylor1993}. Given the 4.4 minute delay between the prompt emission and the start of the LOFAR observations, we are able to probe DM values $\ge 1319$ pc cm$^{-3}$. According to the NE20001 model of free electrons in our Galaxy, the Galactic component of the DM in the line of sight towards GRB 181123B is 56 pc cm$^{-3}$ \citep{cordes2002}. Assuming a relation between DM and redshift \citep[$DM \sim 1200 z$ pc cm$^{-3}$; e.g.][]{ioka2003}, we are able to constrain the prompt emission for redshifts $\gtrsim 1.05$ (assuming zero contribution from their host galaxy and subtracting the contribution of 56 pc cm$^{-3}$ from the Milky Way). 

If the prompt coherent radio emission originates from the same location as the prompt gamma-ray emission, we can constrain the power ratio, $\langle \delta \rangle$:
\begin{eqnarray}
\langle \delta \rangle = \frac{\Phi_{r}}{\Phi_{\gamma}} \label{eqn:powerRat}
\end{eqnarray}
where $\Phi_{r}$ and $\Phi_{\gamma}$ are the radio and gamma-ray bolometric fluences respectively. The model proposed by \cite{usov2000}, in which the radio and gamma-ray emission originate from magnetic reconnection in a strongly magnetized jet, can be constrained using this ratio and it is equivalent to $\langle \delta \rangle \simeq 0.1 \epsilon_{B}$ where $\epsilon_{B}$ is the proportion of energy contained in the magnetic fields. In \cite{rowlinson2019b}, we show for typical GRB gamma-ray spectra and observations at 144 MHz that:
\begin{eqnarray}
\langle \delta \rangle \simeq [4.7 - 7.2]\times10^{9} (1+z)^{0.6} \epsilon^{-0.3}_{B} \frac{\Phi_{\nu}}{\Phi_{\gamma}}
\end{eqnarray}
where $\Phi_{\nu}$ is the fluence limit obtained in the shortest snapshot radio images, corresponding to a 3$\sigma$ limit of $1.8 \pm 0.3 $ Jy s for our 8 second images for GRB 181123B (see Table \ref{table:rms}). The gamma-ray fluence for GRB 181123B was measured by {\it Swift} to be $3.8 \pm 1.2 \times 10^{-7}$ erg cm$^{-2}$ in the 15 -- 350 keV energy band \citep{palmer2018}. 
As stated in Section \ref{sec:otherobs}, the potential redshift of this GRB is $z=1.8$. Thus, we can constrain the fraction of energy stored in the magnetic fields within the relativistic jets of GRB 181123B to be $3 \times 10^{-5} \lesssim \epsilon_{B} \lesssim 2 \times 10^{-4}$ (uncertainties in observed values have been propagated and are included in this range). Therefore, we are able to rule out this model as $\epsilon_{B}$ is at least two orders of magnitude lower than expected for magnetic re-connection models for the prompt emission in GRBs \citep[e.g.][]{beniamini2014}.

Therefore, future observations of high redshift events or with a more rapid slew time would enable us to tightly constrain the \cite{usov2000} model \citep[see also the application of this model to LOFAR follow-up of X-ray flares by][]{starling2020}.

\subsection{Constraints on Fast Radio Bursts}
\label{sec:FRBs}

As outlined in the introduction, a proportion of the FRBs may be associated with the merger of two neutron stars, with the emission originating from mechanisms prior to the merger, during the merger or post merger \citep[see e.g.][and references therein]{rowlinson2019,gourdji2020}. The LOFAR observation of GRB 181123B can be utilised to constrain the emission from dispersed FRBs originating from this source. 

\begin{figure}
\centering
\includegraphics[width=0.48\textwidth]{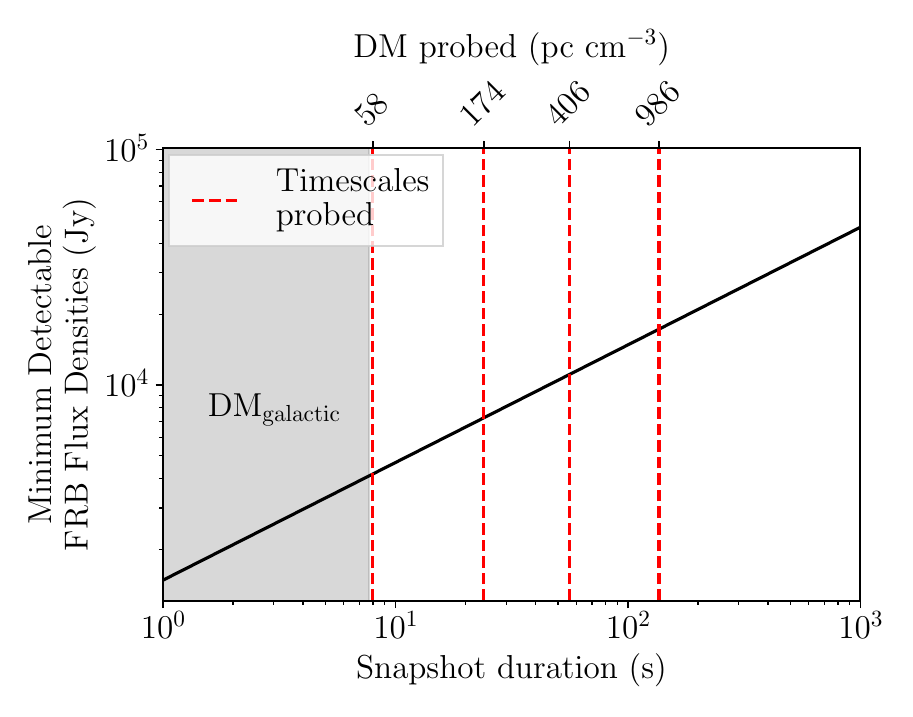}
\caption{The minimum FRB flux density detectable in the LOFAR observations of GRB 181123B. Here we assume that the dispersed signal duration across the frequency band is equal to the snapshot duration. The snapshot durations used in this analysis are shown by the red dashed lines, with their corresponding DM values shown on the top x-axis. The shaded region illustrates the Galactic component of the DM in the sight line towards GRB 181123B.}
\label{fig:dmRange}
\end{figure}

Here, we use the method outlined in \cite{rowlinson2019b} to determine the optimal snapshot time and the associated minimum detectable FRB flux densities at a range of dispersion measures (DMs). For consistency with previous works, the width of an FRB is assumed to be 1 millisecond. The minimum detectable flux densities are obtained using equations 2 and 3 in \cite{rowlinson2019b}, scaled using the 3$\sigma$ upper limit of 126 mJy obtained in the 138 second integrated images. As the rms noise values are not following the typical relationship of $t^{\frac{1}{2}}$ (see Section \ref{sec:Imaging}), this provides conservative minimum detectable flux densities. The optimal snapshot time, for the LOFAR observing band of 120.5 -- 167 MHz, for a given DM value is calculated using equation 4 in \cite{rowlinson2019b}. We assume this GRB occurred within a redshift of 1, corresponding to an IGM contribution to the DM of up to 1000 pc cm$^{-3}$ with a Galactic contribution of 56 pc cm$^{-3}$ \citep[including a contribution of 30 pc cm$^{-3}$ from our Galaxy halo;][]{yamasaki2020,dolag2015}. Then, using a minimum snapshot integration time of 8 seconds, we use 4 snapshot timescales roughly logarithmically spaced to cover this range of 8, 24, 56 and 136 seconds. The minimum detectable FRB flux densities as a function of the snapshot timescale, or DM, are plotted in Figure \ref{fig:dmRange}. In the 8 second snapshot images, we are sensitive to FRBs with flux densities $\gtrsim$400 Jy. No FRBs were detected in the snapshot images of GRB 181123B.

\subsection{Constraints on the magnetar central engine model}
\label{sec:magnetarModel}

As outlined in Section \ref{sec:introduction}, SGRBs like GRB 181123B are believed to originate from the merger of two neutron stars or a neutron star and a black hole. A number of theories have been proposed to produce coherent radio emission between stages prior to and following the merger process \citep[see ][ and references therein]{rowlinson2019b, gourdji2020}. SGRBs exhibiting a plateau phase in their X-ray light curves are consistent with mergers of two neutron stars that combine to form a hyper-massive neutron star with high magnetic fields \citep[hereafter referred to as magnetars;][]{rowlinson2010, rowlinson2013}. If the X-ray plateau phase is followed by a steep decay phase, these are interpreted to be magnetars that are too massive to become stable neutron stars and collapse to form a black hole. As can be seen in Figure \ref{fig:LC}, the X-ray light curve of GRB 181123B shows a plateau phase and a subsequent steep decay phase. Thus GRB 181123B can be explained as a merger of two neutron stars that led to an unstable magnetar that collapsed a few hundred seconds after its formation. Therefore, in this section, we are able to test two of the models outlined by \cite{rowlinson2019} relating to the formation and subsequent collapse of a magnetar.

\subsubsection{Modelling of X-ray light curve}
\label{sec:X-rayModelling}

Assuming a magnetar was produced during GRB 181123B and is powering the plateau phase in the X-ray light curve, we can deduce the key magnetar parameters by fitting the magnetar model to the rest-frame light curve. Following the method outlined in \cite{rowlinson2013,rowlinson2019b}, we take the observer frame X-ray light curve (as shown in Figure \ref{fig:LC}) and convert it to a rest-frame light curve in the 1--10,000 keV energy band. In this conversion, we assume the average SGRB redshift of 0.7 \citep[e.g.][]{rowlinson2013} and use a k-correction \citep{bloom2001}. 

\begin{figure}
\centering
\includegraphics[width=0.48\textwidth]{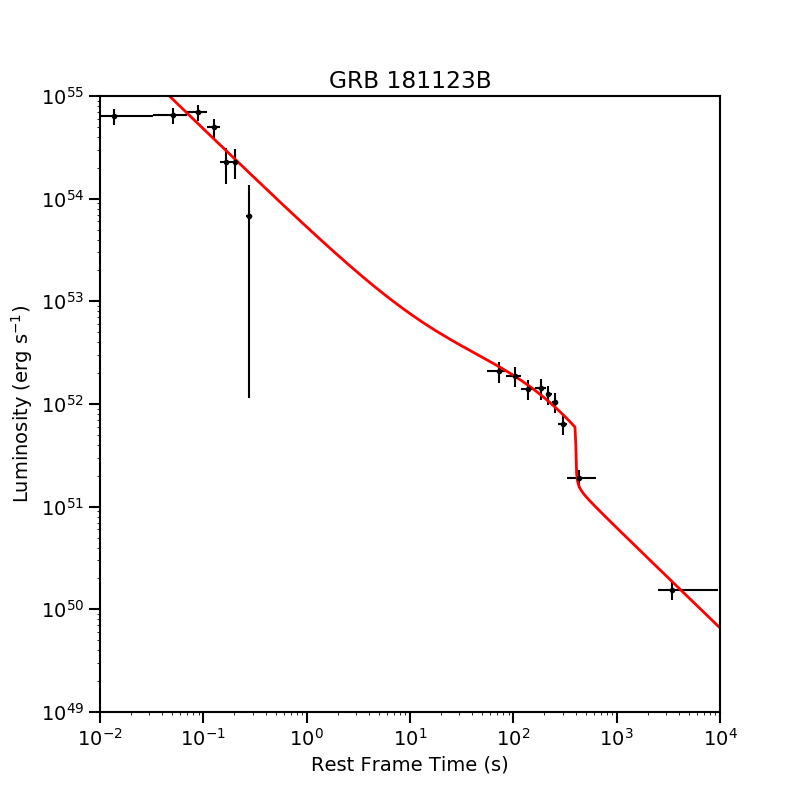}
\caption{This figure shows the rest frame X-ray light curve, assuming a redshift of 0.7. The red line shows the magnetar model fit obtained, corresponding to a magnetic field of $2.4^{+1.3}_{-1.3} \times 10^{14}$ G and spin period of $0.095^{+0.011}_{-0.020}$ ms.}
\label{fig:restframeLC}
\end{figure}

Using the method outlined in \cite{rowlinson2013,rowlinson2019b}, we fitted the magnetar model \citep{zhang2001} to the rest-frame light curve. In the subsequent analysis, we utilise $f \sim 3.45$ (where $f$ is a factor encompassing the beaming angle and efficiency uncertainties) following the analysis of \cite{rowlinson2019b}.


We find that the rest-frame light curve, at a redshift of 0.7, can be fitted with an unstable 1.4M$_{\odot}$ magnetar that collapses at $\sim$400 seconds with a magnetic field of $2.4^{+1.3}_{-1.3} \times 10^{14}$ G and spin period of $0.095^{+0.011}_{-0.020}$ ms (note \cite{sarin2020} also modelled this GRB using a Bayes inference fitting technique and found an earlier collapse time of 250 seconds). In addition to the magnetar component, there is a power-law decay from the prompt gamma-ray emission, with a slope of $\alpha = 0.973^{+0.039}_{-0.040}$. We show this fitted model in Figure \ref{fig:restframeLC}.

For the assumed redshift of 0.7, we find that the fitted magnetar is spinning unphysically fast as it is spinning significantly faster than the spin break-up limit \citep[0.8 ms for a 1.4 M$_{\odot}$ neutron star;][]{lattimer2004}. For a higher mass neutron star of 2.1M$_{\odot}$, as might be expected from a neutron star merger, the spin break-up limit is a lower value \citep[0.55 ms, as calculated using equation 3 of ][ assuming a radius of 10 km]{lattimer2004}. For an unstable 2.1M$_{\odot}$ magnetar we find a magnetic field of $3.6^{+2.0}_{-1.9} \times 10^{14}$ G and spin period of $0.116^{+0.017}_{-0.024}$ ms. However, the fitted spin period at a redshift of 0.7 for this heavier magnetar scenario is still significantly smaller than the 0.55 ms spin break-up period. Therefore, if GRB 181123B is to be explained using the magnetar central engine model, the redshift of the event must be significantly lower than the average SGRB redshift. We can thus constrain the redshift of GRB 181123B by assuming that a magnetar, with spin period less than the spin break-up, was formed. Using the following scalings \citep[from ][]{rowlinson2019,rowlinson2019b}, we can determine the magnetic fields and spin periods as a function of the assumed redshift,
\begin{eqnarray}
B & \propto & \frac{(1+z)}{D_{L}}, \label{eqn:Bpropz} \\
P & \propto & \frac{(1+z)^{\frac{1}{2}}}{D_{L}}, \label{eqn:Ppropz}
\end{eqnarray}
where $B$ is the magnetic field of the magnetar and $P$ is the initial spin period of the magnetar.

\begin{figure}
\centering
\includegraphics[width=0.48\textwidth]{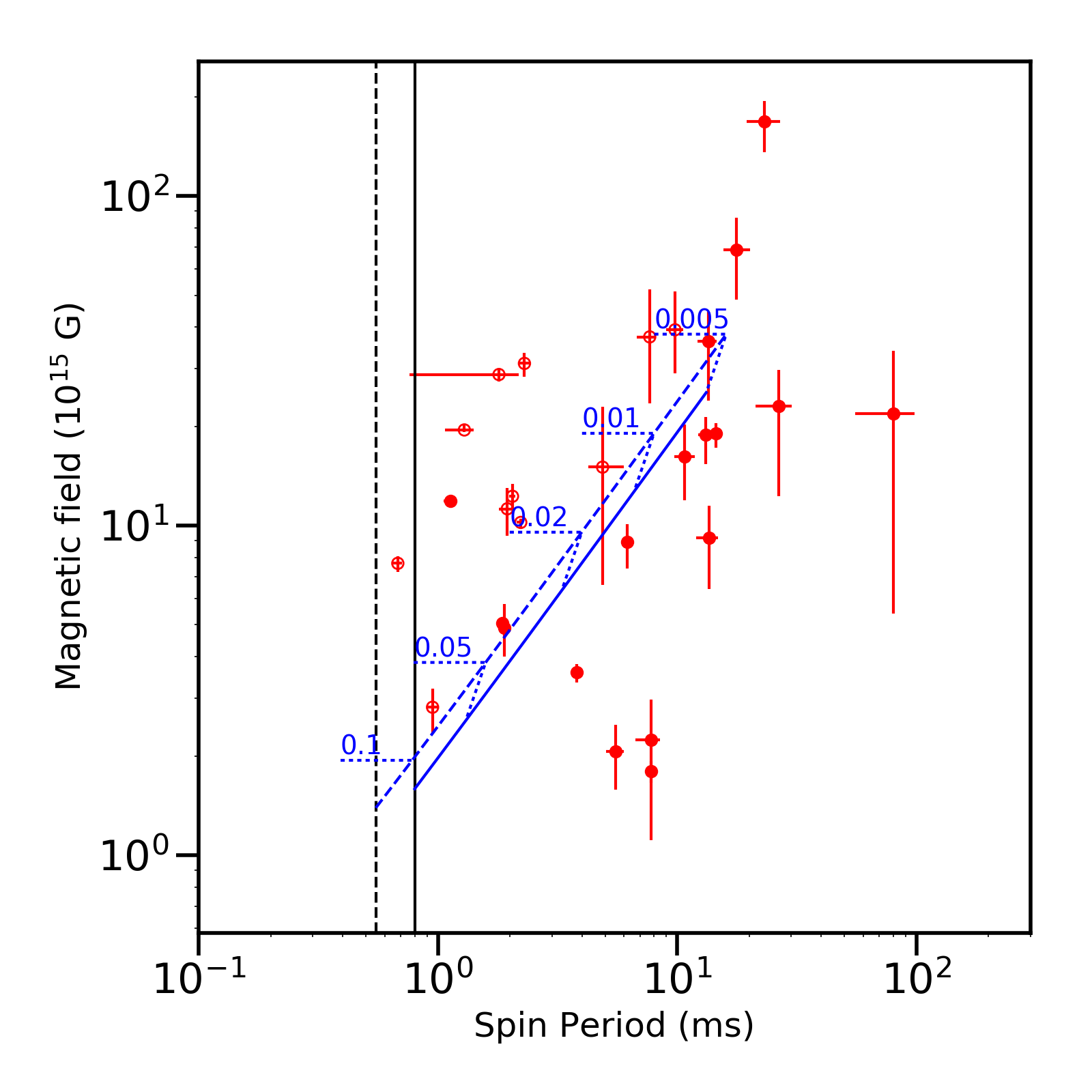}
\caption{This figure shows the magnetic fields and spin periods for the population of SGRBs fitted with the magnetar model \citep[red data points, from ][]{rowlinson2013}. The black vertical lines represent the spin break-up periods for 1.4 M$_{\odot}$ (solid line) and 2.1  M$_{\odot}$ (dashed line) neutron stars. The blue diagonal lines show the solutions for GRB 181123B (1.4 M$_{\odot}$ with a solid line and 2.1  M$_{\odot}$ with a dashed line) at a range of redshifts (some labelled for reference) up to the spin break-up limits. GRB 181123B would need to be at a redshift $<0.08$ to be consistent with forming a 1.4 M$_{\odot}$ magnetar or at a redshift $<0.14$ to be consistent with forming a 2.1  M$_{\odot}$ magnetar.}
\label{fig:BPplot}
\end{figure}

In Figure \ref{fig:BPplot}, we plot the magnetic field and spin period of GRB 181123B, for a range of redshifts from z=0.005 up to the highest redshift attainable before the spin period is faster than that allowed by the spin break up limit. We plot the solutions for both a 1.4 M$_{\odot}$ (solid blue line) and a 2.1 M$_{\odot}$ (dashed blue line) neutron star. For comparison, we also show the population of SGRBs fitted by the magnetar model from \cite{rowlinson2013}. We find the maximum redshifts that GRB 181123B could have occurred at (and still be fitted with the magnetar model) are $z_{\rm max} = 0.08$ and $z_{\rm max} = 0.14$, for the 1.4 M$_{\odot}$ and 2.1 M$_{\odot}$ scenarios respectively. As the fitted model shows that the magnetar collapsed at $\sim$400 seconds, the observations are consistent with a higher mass magnetar being formed. We note that at the putative host redshift of this event, $z = 1.8$, it is impossible to form a magnetar as this model predicts a spin period significantly faster than the spin break up limit.

\subsubsection{Host galaxy constraints}
\label{sec:hostgal}

As noted in Section \ref{sec:otherobs}, a likely host galaxy was observed at a redshift of 1.8 with a probability of chance alignment of 0.44 per cent \citep{paterson2020}. As this host identification has a small, but non negligible, probability of being a chance alignment, we also consider other host galaxy candidates for GRB 181123B. Since compact binary systems can be found at significant offsets from their hosts, due to natal kicks or dynamical processes \citep[e.g.][]{Salvaterra2010,Tunnicliffe2014,Bray2016}, we performed a catalogue search beyond the XRT error circle. 
To take the nearest in separation, we briefly discuss galaxies of interest within 20$''$ of the UVOT-enhanced XRT position. We find one galaxy with redshift only just beyond the $z=0.14$ limit for a magnetar engine following our X-ray analysis in section \ref{sec:X-rayModelling}. SDSS\,J121727.68+143609.2 has a photometric redshift estimate of 0.326$\pm$0.101 \citep[e.g.][]{SDSSDR12} and can be classified as an early-type galaxy when considering the SDSS $ugr$ colours \citep{Strateva2001}. At that redshift and with a separation from the XRT position of 17.6$''$, the implied projected offset is 84 kpc. Two further SDSS galaxies are found at smaller separations of $\sim$7$''$: SDSS\,J121728.29+143546.3 at $z=0.575\pm0.043$ and impact parameter 51 kpc, classified as a red, early-type galaxy and SDSS\,J121728.29+143546.4 (WISEA\,J121728.29+143546.2) which has no redshift but can be classified as a starburst galaxy from its WISE colours, consistent with a late-type classification from the SDSS colours. 

We then carried out a catalogue search using NED and SIMBAD, for all galaxies with known distance $\le 366$ Mpc (corresponding to our X-ray limit of $z\le0.08$ for a 1.4 M$_{\odot}$ magnetar) and projected offset $\le 200$ kpc, using the galaxy-matching method of \citet{Mandhai2019}. 
This results in 5 galaxy candidates. Three are 2MASX sources associated with bright galaxies NGC\,4254, NGC\,4262 and IC\,3065 at distances 13.9--16.6 Mpc and with impact parameters spanning 91-164 kpc. These galaxies have significant separations from the GRB however, of 22--34$'$.
Two are faint galaxies in SDSS-DR12, found at 7$'$ and 2$'$ separations respectively: SDSS\,J121753.41+143228.2, at 76 Mpc distance with impact parameter 156 kpc, which can be classified as an early-type galaxy and SDSS\,J121721.17+143706.6, at 285 Mpc and impact parameter 170 kpc, with a WISE counterpart and classified as a late-type galaxy. These are therefore promising host candidates.

\begin{figure}
\centering
\includegraphics[width=0.48\textwidth]{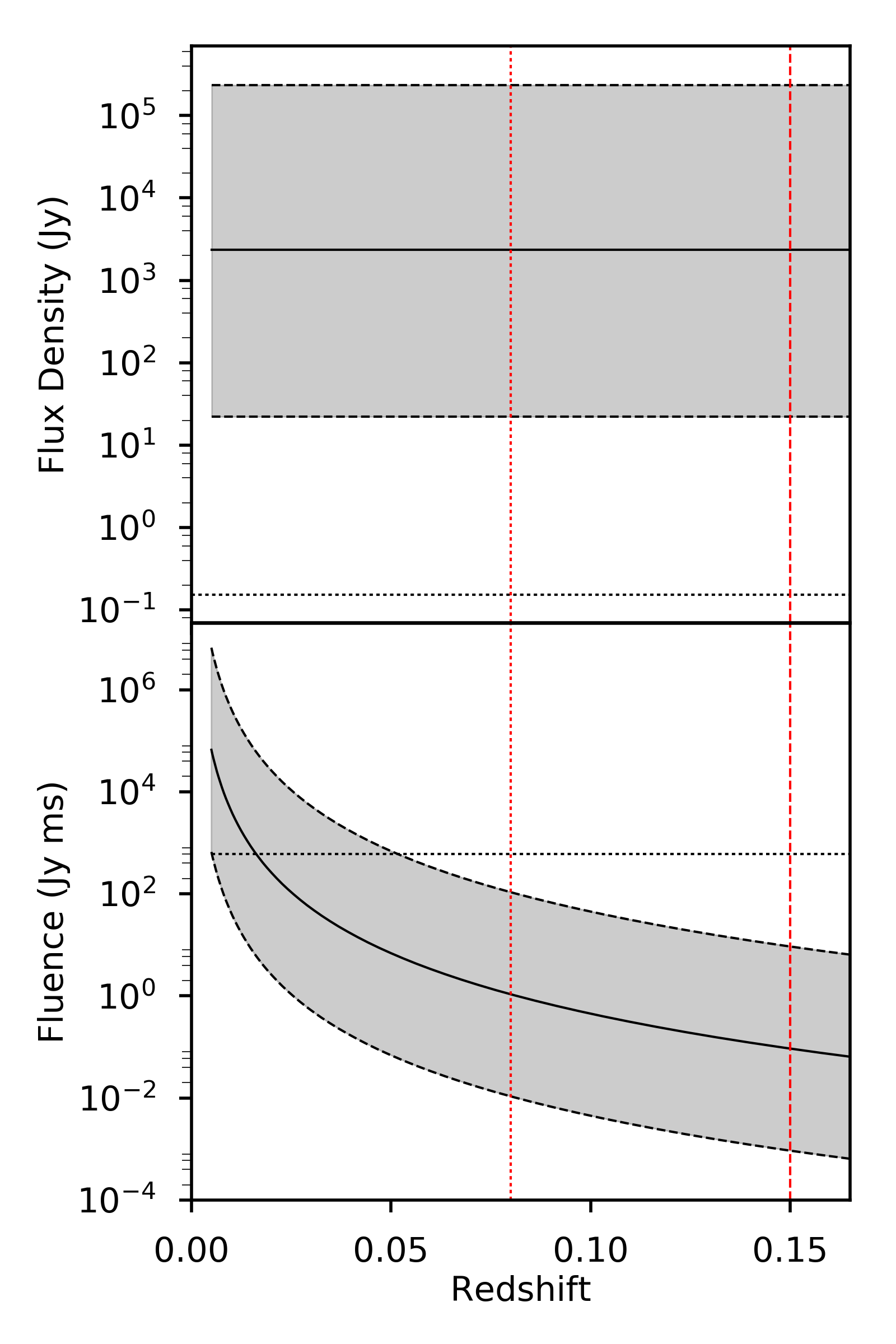}
\caption{In this Figure, we show the predicted emission as a function of redshift for the two coherent emission models considered for GRB 181123B (black solid lines) with their associated 1$\sigma$ uncertainties (black dashed lines and shaded region). The red dotted vertical lines show the maximum redshifts of 0.08 and 0.14, for a 1.4 M$_{\odot}$ magnetar and a 2.1 M$_{\odot}$ magnetar respectively. Top: this shows the predicted flux density for a spinning down magnetar (c.f. Section \ref{sec:pulsar}) assuming a pulsar efficiency of $\epsilon = 10^{-4}$. The observed 3$\sigma$ flux density limit of 153 mJy during the plateau phase is shown as the black dotted horizontal line. Bottom: this shows the predicted fluence for a magnetar collapsing to form a black hole (c.f. Section \ref{sec:ZhangBHmodel}) assuming an efficiency of $10^{-6}$ with the shaded area representing an efficiency range of $10^{-4}$ -- $10^{-8}$. The observed 3$\sigma$ fluence limit of $1.8 \times 10^{3}$ Jy ms for the 8 second images is shown as the black dotted horizontal line.}
\label{fig:modelz}
\end{figure}

\subsubsection{Pulsar like emission}
\label{sec:pulsar}

Assuming a 2.1 M$_{\odot}$ magnetar was formed, we can predict the expected pulsar emission from this source assuming the model proposed by \cite{totani2013} \citep[see also ][]{pshirkov2010} following the method outlined in \cite{rowlinson2019b}. The predictions made in the following section only change slightly for the 1.4 M$_{\odot}$ magnetar and hence the conclusions drawn hold even if the magnetar has a lower mass. \cite{totani2013} showed that the radio flux density of the pulsar like emission produced by the newly formed magnetar is given by
\begin{eqnarray}
F_{\nu} \simeq 8\times10^{7} \nu_{\rm obs}^{-1} \epsilon_{r} D_{\rm lum}^{-2} 
B_{15}^{2} R_6^6 P_{-3}^{-4} ~{\rm Jy} \label{eqn:totaniFlx1}
\end{eqnarray}
where $D_{\rm lum}$ is the luminosity distance in Gpc, $\nu_{obs}$ is the frequency in MHz, $R = 10^{6} R_6$ cm is the radius of the magnetar and $\epsilon_r$ is the efficiency. In this model we assume that the pulsar magnetic field axis is directed towards Earth \citep[see the discussion in ][]{rowlinson2017} and that the magnetar is emitting via dipole radiation \citep[see ][ for further discussion]{lasky2017}. In this analysis, we take the typical pulsar efficiency of $10^{-4}$ \citep[e.g. ][]{taylor1993} but note that this value is poorly known \citep[see ][ for further discussion]{rowlinson2019, rowlinson2019b}. Taking into account the uncertain parameters in the models, namely $f$ and $\epsilon_{r}$, we find that the predicted flux density is given by
\begin{eqnarray}
F_{\nu} = 6.8^{+38}_{-5.9} \times 10^7 ~ \frac{\epsilon_r}{f} ~{\rm Jy}.
\end{eqnarray}
The fraction $f$ can be constrained to be $3.45\pm0.29$ \citep{rowlinson2014,rowlinson2019}. Therefore, given the predicted value of $F_{\nu}$, the allowed range of $f$, and the observed LOFAR upper limit of the flux density at the position of GRB 181123B in the 400 seconds integrated observation of 153 mJy, we find that $\epsilon_r \le 6\times 10^{-8}$. Therefore, as no emission was detected, it is shown that GRB 181123B either did not form a magnetar (consistent with the expectation if associated with the candidate host galaxy at $z=1.8$), the surrounding medium is effectively blocking the emission, the beaming of the radio emission is different to that of the X-ray emission or the model proposed by \cite{totani2013} is not correct. In \cite{anderson2020}, they consider the same model for the short GRB 180805A, however for that event the emission was predicted to be too faint to be observable due to having a lower energy magnetar (from fitting the X-ray light curve its spin was slower and the magnetic field was lower than that fitted for GRB 181123B). More observations of other GRBs will be required to increase the statistical significance of this non-detection and to explore the observed range of magnetic fields and spin periods.

\subsubsection{Emission associated with collapse to black hole}
\label{sec:ZhangBHmodel}
Assuming a magnetar was formed during SGRB 181123B, the X-ray fitting suggests it was an unstable neutron star that collapsed to form a black hole at $\sim$400 seconds abruptly ending the plateau phase (see Section \ref{sec:X-rayModelling}). This collapse is thought to be accompanied by a brief flash of coherent radio emission as magnetic reconnection occurs within the pulsar magnetosphere \citep{falcke2014, zhang2014}. As shown in \cite{rowlinson2019b}, the observed flux density, $f$, at a given observing frequency, $\nu_{obs}$, can be described as:
\begin{eqnarray}
f_{\nu} = - \frac{10^{-23} \epsilon E_{B}}{4 \pi D_{lum}^2 \tau}(\alpha+1) \nu_{p}^{-(\alpha+1)} \frac{\nu^{\alpha}_{obs}}{(1+z)} \frac{\tau}{t_{int}} ~{\rm Jy}, \label{eqn:collapseFlux} 
\end{eqnarray}
where $E_{B}$ is the amount of energy available in the magnetic field of the neutron star in erg (given by $E_{b} = 1.7 \times 10^{47} B^{2}_{15} R^{3}_{6} ~erg$), $\epsilon$ is the fraction expected to be converted into coherent radio emission, $t_{int}$ is the intrinsic duration of the emission in seconds \citep[assumed to be 1 millisecond in this analysis;][]{falcke2014}, $\alpha$ is the spectral index of the radio emission, $\nu_{p}$ is the plasma frequency given by $\nu_{p} \simeq 9 n_{e} ~kHz$, $n_e$ is the number density of electrons in $cm^{-3}$ and $t_{int}$ is the integration time of the image in seconds.

For GRB 181123B, we assume a spectral slope of the coherent radio emission to be $\alpha = -2$, the efficiency to be $\epsilon = 10^{-6}$ and take the magnetar parameters from Section \ref{sec:X-rayModelling}. In Figure \ref{fig:modelz}, we plot the predicted fluence for this event (with the upper and lower bounds being for $\epsilon = 10^{-4}$ and $10^{-8}$ respectively) and over plot the fluence limit attained in our shortest snapshot images (a flux limit of 76 mJy in an 8 second image corresponds to $1.8 \times 10^{3}$ Jy ms). From this analysis, we would only have a chance of detecting this emission if GRB 181123B occurred at a redshift $\lesssim 0.05$ and the efficiency is of order $10^{-4}$. We note that if the assumed spectral index was larger than $\alpha = -2$, this emission would be easily detectable by our LOFAR observations.

Our non-detection can be interpreted in a number of ways. Firstly, the efficiency and spectral index of the emission is low and hence it is undetectable in our images. Making shorter duration images and/or conducting image plane de-dispersion \citep[as in;][]{anderson2020} will increase the chance of detection. Secondly, the merger was expected to be at a higher redshift than 0.05 and hence too distant to be detected. Thirdly, the interpretation that the X-ray light curve shows the formation and collapse of a magnetar is incorrect. Fourthly, the emission is beamed away from us or is unable to propagate through the surrounding medium. As found in Section \ref{sec:pulsar}, more observations of other GRBs are required to determine which is the most likely interpretation for this non-detection.

\section{Conclusions}
\label{sec:conclusions}

In this study, we have found no evidence for low frequency, coherent radio emission originating from SGRB 181123B. We searched for persistent emission during the plateau phase and short duration radio flares throughout the full observation. We have compared this non detection to theoretical models to place deep and constraining limits. Assuming the putative host galaxy redshift, we are able to rule out theoretical models as outlined below. As there is still some uncertainty given the possibility of a chance alignment with that galaxy, we also draw conclusions assuming that the redshift is unknown.

Assuming that GRB 181123B was at a redshift greater than 1.05, we are able to show that our non detection implies that the fraction of energy contained within the magnetic fields in the relativistic jets is an order of magnitude lower than expected for magnetic reconnection models for GRB prompt emission. This limit decreases to being two orders of magnitude lower than the models for the case of the redshift of the likely host galaxy. More rapid slew times for radio telescopes are required to be able to constrain this model for lower redshifts. \cite{starling2020} also aims to constrain this model at low redshift for X-ray flares, which are believed to be from the same emission mechanism as the prompt gamma-ray emission, by exploiting the observed delay between triggering on the prompt emission and the X-ray flares enabling simultaneous radio observations.

The X-ray light curve of GRB 181123B shows evidence of on-going energy injection during the first few hundred seconds before stopping abruptly and resuming a power law decay associated with the afterglow phase. We interpret this energy injection as resulting from a newborn magnetar, formed via the merger of two neutron stars, which collapses to form a black hole at $\sim$400 seconds. By fitting the X-ray light curve, we were able to constrain the magnetar parameters required to test coherent radio emission models. We find that the magnetar parameters are unphysical for redshifts $>$0.14 for a 2.1 M$_{\odot}$ neutron star ($>0.08$ for a 1.4 M$_{\odot}$ neutron star). Thus, if this model is correct, we would expect the event to occur at a low redshift. We find there are catalogued galaxies that would be consistent with this interpretation, all offset from the GRB localisation implying significant kicks. However, at the proposed redshift of the likely host galaxy, this model and it's associated coherent radio emission is ruled out.

The first model tested was that of pulsar like emission from the newly formed magnetar during the energy injection phase \citep[e.g.][]{totani2013}. We find that the predicted emission is $\sim$4 orders of magnitude brighter than the upper limited obtained assuming an efficiency of $10^{-4}$ expected for standard pulsars. We can constrain the efficiency of conversion of rotational energy into coherent radio emission for the newborn neutron star to be $\le 6 \times 10^{-8}$. Explanations for this non-detection that require further study include absorption of the emission by the surrounding medium or the beaming of the X-ray emission is different to that of the radio emission.

The second model tested predicted a short flash of coherent radio emission when the magnetar collapses to form a black hole and magnetic reconnection of the field lines occurs \citep{falcke2014,zhang2014}. We find this emission would only be detectable for events at redshifts $\le 0.05$ for an assumed efficiency of $10^{-6}$ and a radio spectral index of -2. Therefore, our non-detection is consistent with this model.

Analysis of GRB 181123B, shows the ability of the current generation of radio telescopes to extensively test these emission model theories. With observations of more neutron star binary mergers (via triggering on cosmological SGRBs or low redshift gravitational wave events) and careful modelling, we will either detect this emission, be able to show the emission models are incorrect or that the surrounding medium is opaque to coherent radio emission.

\section*{Data Availability}

The data and scripts underlying this article are available in Zenodo, at \url{https://doi.org/10.5281/zenodo.3957613}. The LOFAR data used are available in the LOFAR Long Term Archive at \url{https://lta.lofar.eu} under the Cycle 11 project title LC11{\_}002 with SAS Id numbers 689616 (target data) and 689622 (calibrator data). The {\it Swift} data used are available from the UK {\it Swift} Science Data Centre at the University of Leicester at \url{https://www.swift.ac.uk/index.php} under GRB 181123B.  

\section*{Acknowledgements}
We thank the LOFAR Radio Observatory for implementing the new rapid response mode and for supporting our observations. 

RAMJW acknowledges funding from the ERC Advanced Investigator grant no. 247295. RLCS acknowledges support from the ASTRON Helena Kluyver Visitor Programme and from STFC. 
SM is supported by a PhD studentship awarded by the College of Science and Engineering at the University of Leicester.
GEA is the recipient of an Australian Research Council Discovery Early Career Researcher Award (project number DE180100346) funded by the Australian Government.

This paper is based (in part) on data obtained with the International LOFAR Telescope (ILT) under project code LC10\_012. LOFAR \citep{vanhaarlem2013} is the Low Frequency Array designed and constructed by ASTRON. It has observing, data processing, and data storage facilities in several countries, that are owned by various parties (each with their own funding sources), and that are collectively operated by the ILT foundation under a joint scientific policy. The ILT resources have benefited from the following recent major funding sources: CNRS-INSU, Observatoire de Paris and Universit{\'e} d'Orl{\`e}ans, France; BMBF, MIWF-NRW, MPG, Germany; Science Foundation Ireland (SFI), Department of Business, Enterprise and Innovation (DBEI), Ireland; NWO, The Netherlands; The Science and Technology Facilities Council, UK; Ministry of Science and Higher Education, Poland.

This work made use of data supplied by the UK {\it Swift} Science Data Centre at the University of Leicester and the {\it Swift} satellite. {\it Swift}, launched in November 2004, is a NASA mission in partnership with the Italian Space Agency and the UK Space Agency. {\it Swift} is managed by NASA Goddard. Penn State University controls science and flight operations from the Mission Operations Center in University Park, Pennsylvania. Los Alamos National Laboratory provides gamma-ray imaging analysis.

This research has made use of the NASA/IPAC Extragalactic Database (NED), which is operated by the Jet Propulsion Laboratory, California Institute of Technology, under contract with the National Aeronautics and Space Administration. This research has made use of the SIMBAD database, operated at CDS, Strasbourg, France



\bibliographystyle{mnras}
\bibliography{bibliography.bib} 


\bsp	
\label{lastpage}
\end{document}